\documentclass[12pt]{iopart}
\usepackage{graphicx}

\begin{document}
\title{Dynamic Phase Transition in Prisoner's Dilemma on a Lattice with Stochastic Modifications.}
\author{M Ali Saif}
\address{Centre for Modeling and Simulation\\
University of Pune\\
Ganeshkhind, Pune, 411 007,\\
INDIA\\}
\ead{ali@cms.unipune.ernet.in}
\author{Prashant M Gade}
\address{Department of Physics\\
Rashtrasant Tukdoji Maharaj Nagpur University\\ 
Campus, Nagpur, 440 033\\
INDIA\\}
\ead{prashant.m.gade@gmail.com}

\begin{abstract}
We present a detailed study of prisoner's dilemma game with stochastic
modifications on a two-dimensional lattice, in presence
of evolutionary dynamics. By very nature of the rules, 
the cooperators have incentive to cheat 
and the fear of being cheated.
They may cheat even when not dictated by evolutionary dynamics.
We consider two variants here. 
In either case, the agents do mimic the action 
(cooperation or defection) in the previous timestep
of the most successful agent in the neighborhood. 
But over and above this,  
the fraction $p$  of cooperators spontaneously change their strategy
to pure defector at every time step
in the first variant. In the second variant, there are no pure cooperators.
All cooperators
keep defecting with probability $p$ at every time-step.
In both cases, the system 
switches from coexistence state to an  all-defector
state for higher values of $p$. We show that the transition between these
states unambiguously belongs to directed percolation
universality class in $2+1$ dimension. We also study the local persistence.
The persistence exponents obtained are higher 
than ones obtained in previous studies underlining their dependence
on details of dynamics.
\end{abstract}
\maketitle

\tableofcontents

\section{Introduction} 
Cooperation is observed at many levels of 
biological organization. 
The evolution of cooperation in these systems 
has been a subject of extensive debate and studies\cite{nowak00}. 
Primarily, five different mechanisms have been proposed 
to explain how natural selection lead to cooperative behavior. 
They are kin selection, group selection, direct or indirect 
reciprocity and network reciprocity \cite{nowak0}. 
Kin selection explains the cooperation  between 
genetically close organisms as a tendency
to help reproductive success of the relatives even at a cost
to themselves \cite{hamilton}. Cooperation may evolve not only on 
individuals level but also in groups. Thus, a group of 
cooperators are more likely to survive and grow than group of 
defectors. However, some authors believe that, the kin 
selection models are not different from the group 
selection models \cite{lehmann}.   
Cooperation is also observed between
organisms who are not genetically close.
Reciprocal altruism is a possible mechanism to explain the cooperation 
between such agents \cite{trivers,axelrod2}. 
Emergence of sustained cooperation  when  agents
have an incentive to cheat as well as 
tension of being defected \cite{macy}, 
has been a topic of extensive investigation.
In this case, the benefit is extended to another organism 
in the hope that it will be reciprocated in future and this
strategy is reversed if the act is not reciprocated.
Sustaining such strategy is more likely in an iterated or spatial 
game theoretical model.
We would like to mention that the cooperation
is not always direct. Sometimes we help strangers, and there is no 
possibility for direct reciprocation. 
We would like to also mention that
altruism is still an open problem. 

Among various attempts at constructing a theory of 
cooperation, game theoretical models have played 
an important role \cite{szabo0}.  In particular, Prisoner's Dilemma 
(PD) has emerged as a paradigm for the explanation 
of cooperative behavior among selfish individuals \cite{axelrod}. 
This kind of cooperative behavior observed in real life in 
systems ranging from biological  to economic 
and social systems \cite{turner1}. PD has now become
a standard model to explain cooperation in these 
systems \cite{axelrod2,turner2,nowak,doebeli}. 
In its original form, PD 
describes the pairwise interaction between two players. The player
either cooperates($C$) or defects($D$) 
at any confrontation. If both players choose to 
cooperate (defect), they get a pay-off of magnitude $R$ ($P$) each; 
if one($D$)  chooses to defect,  while the other($C$) chooses to
cooperate, the defector  gets the biggest 
pay-off $T$, while the other gets $S$.
For $T>R>P>S$ and $2R>T+S$, 
total reward for both players 
is higher if they cooperate. However, an individual  has
better payoff if he defects while the other player cooperates.
Thus the best choice for any player is to defect 
irrespective of the opponent's choice if the game played for one round.
However, on a two-dimensional lattice, it was found that a  
fraction of players keep cooperating with their neighbors with repeated
interaction.
{\em{Thus, with repeated interactions and spatial structure, it was found that 
it is possible to have mixed state where 
clusters of cooperators coexist with defectors.}} We must mention that recently 
other spatial 
structures have also received a fair share of attention
\cite{abramson,szabo1,vukov1,
kim,vukov2,cesr}.

In the  context of ecology,
Nowak and May simulated PD game with choice of 
parameters $R=1.0$, $T=b$ $(1.0<b<2.0)$ and $S=P=0.0$.
(Some authors called this game 'weak dilemma' when $S=0$ 
\cite{szabo0,alonso}. However, 
Nowak and May \cite{nowak1} found that their qualitative results
does not change when $S<0$, at least for small absolute values of $S$,
{\it {i.e.}}  $|S|<<1$. Hence, we work with $S=0$ in this paper. We have
studied the case when $S=-0.01,-0.1$ and $S=-0.5$ to demonstrate  that our main results
do not change for $S<0$.)
They believe that, with this choice of parameters most 
of the interesting behavior is reproduced. They studied 
PD on a two dimensional array with synchronous updating and explored 
the asymptotic behavior  for various values 
of the parameter $b$. Players interact with their 
local neighbors through simple deterministic rules and 
have no memory of past \cite{nowak1,nowak2}.

This explanation was debated and robustness of the conclusions was
studied under several perturbations of the model. 
Mukherji {\it{et al.}}
as well as Huberman and Glance studied the system under introduction
of asynchronicity \cite{mukherji,huberman,ali-1st}. 
Mukherji et al. investigated  
if cooperation can survive in the spatial PD in the presence of noise
in general. They considered  some more stochastic variants of this system.  
Other modifications by Mukherji {\it {et al.}} were random introduction
of cooperators and defectors at any site and  
spontaneous conversion of cooperators
into defectors with some probability \cite{mukherji}. 
Nowak {\it{et al.}} replied stating
that, their results are robust with respect to these modifications.
If one studies the
entire parameter regime, cooperation is found to persist in the system
even in presence of high values of noise \cite{nowak3}.
We will make a detaied study of one of the  cases studied by 
Mukherji {\it {et al.}}
in which cooperators turn into defector spontaneously with probability
$p$. We call this model as model stochastic prisoner's dilemma (permanent)
(abbreviated as SPD(P)).
Mukherji {\it{et al.}} simulated SPD(P) on a 
$100\times100$ lattice, for $500$ generations with an initial condition 
of $90\%$ cooperators. They found that, the density of 
cooperators quickly decreases with $p$ and above certain value of $p$
all agents become defor increase up to point where all 
players become defector\cite{mukherji}. 
This variant was criticized by Nowak {\it{et al.}} as `this 
assumption is well chosen for attempting to eliminate cooperators'
\cite{nowak3}.
Hence, we will study one more variant,  stochastic prisoner's 
dilemma (temporary) (SPD(T)).
 In this model, each cooperator turns into
defector {\em{temporarily}} with probability $p$ and returns to 
being cooperator at the next time step. In both models, each 
agent imitate the best (unconditional imitation) strategy of neighboring 
agents in last time-step.

In this paper, we make an extensive study of SPD(P) and SPD(T)
from the viewpoint of dynamic phase transition. In both cases, 
only one absorbing 
state is possible, namely, in which all players choose to defect at all 
times. Coexistence of $D$ and $C$ is considered as active phase 
or fluctuating phase.
For large values of noise $p$, we observe a transition from coexistence
state to an all-defector state. 
This kind of phase transition to absorbing state
has  attracted much attention 
recently \cite{marro,odor,lubeck2,hinrichsen}. 
We can study this phase transition borrowing tools used extensively 
for studies in equilibrium systems,
We find whether or not transition is continuous and 
find the several critical exponents. We also find the scaling functions which
give a better idea of universality.
At the basic level, the critical exponents allow us to classify the  
system in different universality classes.
The concept of the universality is one of the 
most important concepts in study of 
phase transitions. It allows us to group different systems to 
small number of classes and lets us know the essential and not so essential
details of the systems.  It is generically believed that, 
all the continuous phase transitions from fluctuating phase 
to a single absorbing state are in the universality class of 
directed percolation (DP) \cite{hinrichsen}. 
(Most system with multiple 
absorbing state also fall in the DP class \cite{marques}.) 
However, under some additional 
conditions, the systems with absorbing state may  not fall under the DP 
class. The other well known universality class for such systems   are parity 
conserving class \cite{takayasu,jensen}, the pair contact process with 
diffusion \cite{odor2,hinrichsen2}, the conserved lattice gas \cite{park,lee1} 
and Manna class \cite{manna}. 
These systems has 
been studied extensively in past two decades \cite{lubeck2,hinrichsen}.  
The SPD(P) and SPD(T) 
have an unique absorbing state, no obvious 
conservation laws, so we expect them to be  in the DP 
universality class.  
We must mention that, phase transition to DP class has been observed 
in the PD game in a previous studies \cite{szabo3,chiappin,hauert,guan}. 
However, there are a few differences between these works and the present one. 
The updating rule in these works was different from the updating rule
suggested by Nowak and May.  Most of earlier
studies computed only the static exponent $\beta$. In this work, we
have made exhaustive and systematic simulations and found
all the three independent exponents of DP class. (In fact, we also
find the fourth exponent and explicitly demonstrate time reversal symmetry.)
Qualitatively, we have
only one possible absorbing state in this system unlike previous
studies where there are two  absorbing states are possible. Thus there
is a stronger ground for Janssen and Grassberger's, \cite{hinrichsen} conjecture to hold in
this case. 
In the other hand, these transitions could be
discontinuous. In fact, experimentally discontinuous transitions are
observed often and unfortunately
there are no clear thumb rules on when the transition is 
continuous and when it is not except in cases where mean field theory is
applicable \cite{Bidaux}. However, the transition in our case is 
clearly continuous.

In this paper, we make a complete study of transition to absorbing state 
of the above two models and we find that they are
indeed in DP class. Furthermore, we study persistence in these
systems. Recently, there have been several studies on persistence 
in dynamic phase transitions. They lead to nontrivial
exponents which 
have no obvious relation with other  critical exponents in the system
since persistence takes into account time correlations of arbitrary order.
We study local persistence in
these two models  and  determine the value of the 
local persistence exponent. Our studies further support the
fact that the  conjecture  of superuniversality
of the exponent \cite{hinrichsen} made in initial studies is not true. 
(Superuniversality means having
an exponent which is independent of dimension.) 
In fact, the exponent is not even 
universal in the sense that different models in same universality class
in same dimension yield widely different exponents. The
exponents obtained by us are  much larger 
than ones obtained in previous studies for some other
systems belonging to DP class.  With systematic numerical studies,
we will demonstrate that the two variants studied by
us are unambiguously in DP universality class. However, they
show different persistence exponents. This is not entirely 
unexpected since then persistence exponent probes a full non-Markovian
evolution of the system and is one of the least universal exponents.

The dependence of persistence
exponent on detailed dynamics of the system has been observed previously in other
systems such as spin systems \cite{cueille}.
Our results further demonstrate that,  
the persistence exponent is very much dependent on the dynamics
and having the same exponent in two different systems
could simply be a coincidence.

\section{Definition of Models and Simulation Results}
We consider evolutionary PD game on 
the two dimensional lattice of size $L$. Each  lattice  site
can take only two values $s=0$ (defector) or $s=1$ (cooperator). 
We fix boundary condition and players have no memory of 
the past. The parameters are chosen to be $R=1$ , $S=P=0$. We study
the model under variation of  $T=b$. 
Each player $(i,j)$ interacts with his eight nearest neighbors 
(Moore neighbors) and himself. The total pay-off of any player 
$p_{(i,j)}(t)$ 
is the sum of the pay-offs from all nine interactions (with neighbors 
and self). In each Monte Carlo step, each player is allowed to 
update his strategy by adopting the strategy of the 
most successful neighbor.  In  SPD(P), after each Monte Carlo step, each cooperator may 
choose to change his state to defector state with probability $p$. In SPD(T), each cooperator  
defects with his neighbors with probability $p$, but unlike SPD(P),  
returns to cooperator status in the next time step even if temporary 
defection has delivered him a good payoff. If his payoff
is lesser than any agent in the neighborhood 
(except himself), he mimics that neighbor's strategy {\em{in previous 
timestep}}. 
If the successful neighbor has cooperated (defected) in previous time step, he 
becomes cooperator (defector). In both cases, for $p=0$ we recover the PD on the two 
dimensional lattice where agents are either pure cooperators 
or pure defectors. Furthermore, in both cases, the only 
possible absorbing state is an all-defector state for any value of $p\neq 0$
(We have also checked that, both models still show DP phase
transition if we make $S<0$ at least when $|S|<<0$.)

The order parameter in this case is the density of cooperators(active) 
sites $\rho(t)=\left\langle 1/N\sum_i s_i(t)\right\rangle$ where $N$ is the 
total number of a lattice sites and $\left\langle ...\right\rangle$ 
denotes to the ensemble average. Clearly, this parameter has a nonzero
value in the mixed state while it is zero in an all-defector state.
For both models SPD(P) and SPD(T),
we plot a phase diagram for the  asymptotic states in
the phase space $(p,b)$. We explore the range  
$0.0<p<1.0$ and $1.0<b<2.0$ for the two 
parameters. The corresponding phase diagram is shown in Fig.1. As 
we can see, the system has two different phase all defector phase 
(absorbing phase) and mixed phase (active phase). For fixed value
of $b$, we vary $p$ and study the nature of phase transition in
the system. 

\begin{figure}
 \includegraphics[width=70mm,height=60mm]{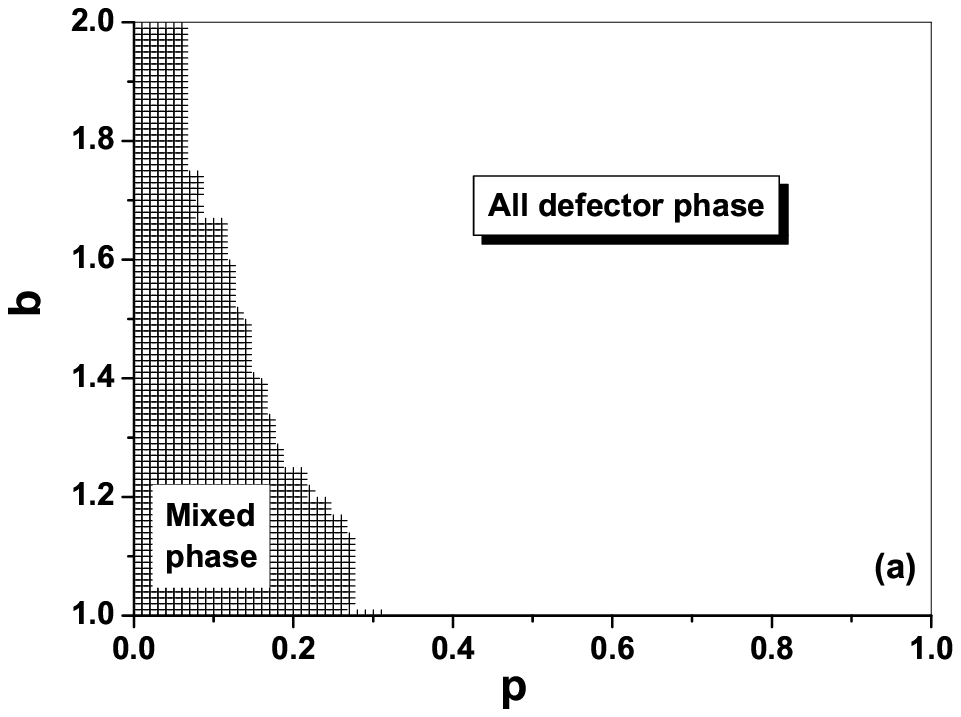}
 \includegraphics[width=70mm,height=60mm]{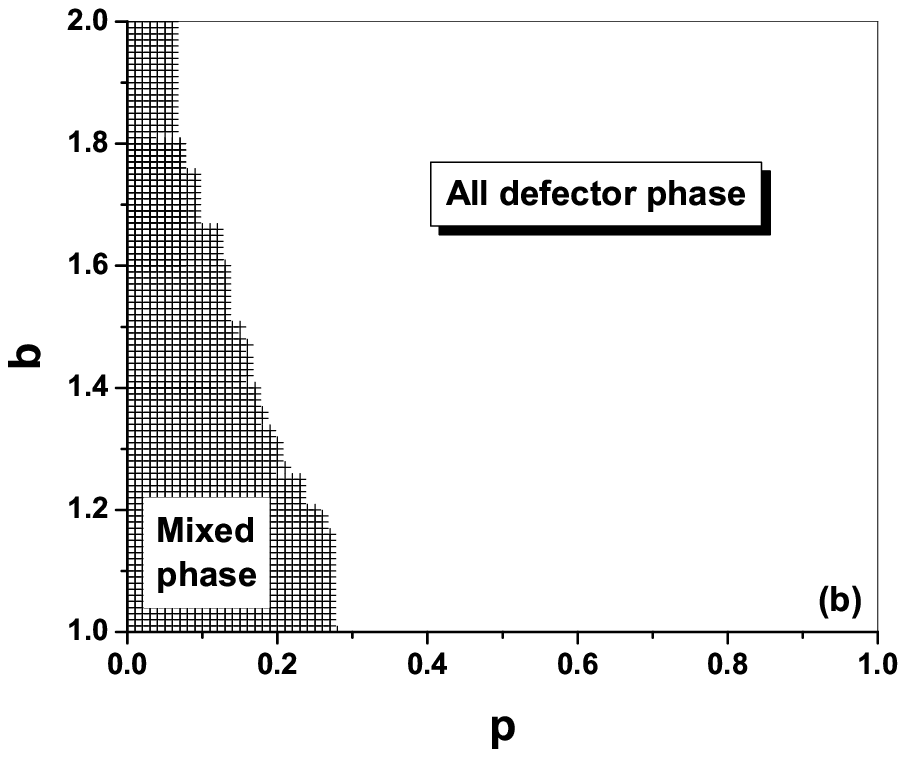}
 \caption{\label{fig1}Schematic phase diagram of (a) SPD(P) and (b) SPD(T) when $S=0$. The shaded 
area corresponds to an active phase and white area corresponds
to an all-defector phase.
We used lattice size $L=60$ and averaged over $100$ different 
initial samples after discarding $1000$ time-steps }
\end{figure}

First, we state results for SPD(P) model.
We simulate the system on large enough lattice, {\it{i.e.}}
on a lattice of size $L=200$. We estimate the
steady state of the order parameter $\rho_{sat}$ by simulating
system for very long time and confirming  that the order parameter
has reached its steady state.  This procedure is carried out for 
various values of $p$.  Initial condition consists of 
$30\%$ defectors and $70\%$ cooperators distributed randomly on the 
lattice sites. For every value of $p$, we average over $100$ different 
initial configuration after discarding $10^5$ timesteps 
near the critical point and $10^4$ timesteps far from the critical 
point. In Fig. 2, we plot the average density of active sites $\rho_{sat}$ as 
function of the control parameter $p$. 
It is clear that, the stationary density 
of the active sites varies continuously with  $p$. The 
system crosses from absorbing phase to active phase at the critical point 
$p_c$. To determine the critical point $p_c$ accurately, we 
use different lattice sizes up to $L=512$. In all cases , we calculated 
the value of $\rho_{sat}$ near the critical point after discarding $10^6$ Monte 
Carlo steps. The best estimate of the value of a critical point 
in thermodynamic limit seems to
converge to $p_c=0.2708 \pm 0.0005$.

\begin{figure}
 \includegraphics[width=70mm,height=60mm]{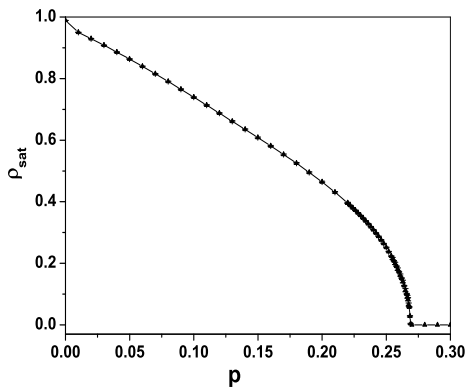}
 \caption{\label{fig2}The steady state of the density of active sites at 
various value of parameter $p$ for SPD(P) at $b=1.05$.}
 \end{figure}

In order to confirm that, phase transition in these models 
is in DP class, we numerically determine the values of all the critical 
exponents. The absorbing phase transitions are characterized by four 
independent critical exponent $\beta$, $\acute{\beta}$, 
$\nu_\bot$ and $\nu_\|$. However, it is well known that,
DP class displays a symmetry known as rapidity reversal symmetry.
This implies that $\beta=\acute{\beta}$. (We must mention that
this statement is easily proven  for directed bond percolation. \cite{hnr}
It is not obvious
 for several other models.   
Hence we explicitly 
checked this symmetry by computing all the exponents.) 
Thus, DP is
characterized only by three critical exponent instead of
four. All the other exponents can be expressed in terms
of these exponents. The so called dynamic exponent $z$ is
given by $z={\frac{\nu_\|}{\nu_\bot}}$. However, the
exponents $\delta$, $\alpha$ and $\theta$ are given by
$\delta={\frac{\acute{\beta}}{\nu_\|}}$, $\alpha={\frac{\beta}{\nu_|}}$
and $\theta=d/z-2\delta$ (for more details see
\cite{lubeck2,hinrichsen} and the references therein).
We would like to emphasize that, in this work we have verified
the equality of $\beta$ and $\beta'$ by computing survival probability
$P(t)$ as
well as density of active sites
$\rho(t)$ independently. If $\beta=\beta'$, both would decay
with the same exponent. By finding effective exponents for $P(t)$ and
$\rho(t)$,  and showing that they are equal, we have verified this symmetry 
for these models (See Fig. 4 and 7.)
 
It is known that, for continuous phase transition, the 
stationary value of 
order parameter $\rho_{sat}$ vanishes as the control parameter $p$  
approaches a critical 
value $p_c$  asymptotically according to a power-law as follows:
  
\begin{equation}
\rho_{sat} \sim (p_c-p)^{\beta}
\end{equation}
The value of exponent $\beta$ can be
found by
plotting the value of $\rho_{sat}$  as a function of 
$(p-p_c)$ on a  logarithmic 
scale Fig. 3. The power-law behavior is clear  and the 
best-fit value of the critical exponent is found to be $\beta=0.57\pm 0.01$
 which matches very well with value of $\beta=0.58$ in the 
DP class \cite{hinrichsen}.  The compatibility of this exponent with the DP in 
the $2+1$ dimension, 
leads us to conjecture that, SPD(P) belongs to the directed 
percolation universality class in $2+1$ dimension.

\begin{figure}
 \includegraphics[width=70mm,height=60mm]{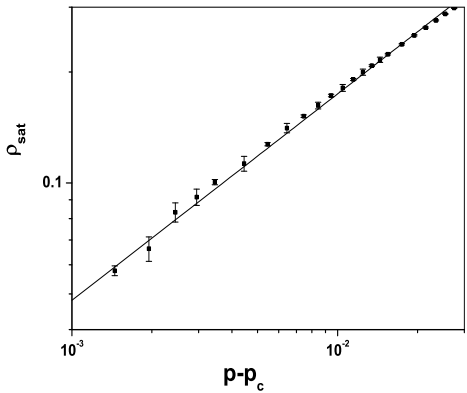}
\caption{\label{fig3}Stationary density of active sites $\rho_{sat}$ 
is plotted as a
function of the distance to the phase transition in log-log scale
for SPD(P). The linear 
fit accurately fits to the numerically obtained data with the exponent 
$\beta=0.57 \pm0.01$.}
 \end{figure}

To be sure about the universality class,
we extract further critical exponents. Finding the nature
of phase transitions is an `asymptotic' game, in the sense that
we need to make conjectures about asymptotic behavior of
thermodynamic system by systematically simulating systems of finite size
for finite time. Fortunately some of the information 
about nature of transition can be inferred from short
time dynamics. Thus it is a simplest numerical method which 
allows us to estimate some of the critical exponents.  
We start Monte Carlo simulation with a fully occupied lattice 
\cite{hinrichsen,silva,lubeck}.  
 At the critical point $p_c$, the 
order parameter $ \rho(t)$  decays asymptotically according to a 
power-law
\begin{equation}
\rho(t)\sim t^{-\delta}
\end{equation}
In Fig. 4a and 4b, we plot $\rho(t)$ as a function of time $t$ in a logarithmic
scale for both the models.  
At the critical point, the order parameter $\rho(t)$ shows
a power-law decay.  The best fit 
of the critical exponent is $\delta=0.456 \pm0.001$ for SPD(P) and $\delta=0.434 \pm 0.002$ for SPD(T), which again 
is in good agreement with  the value 
$\delta= 0.451$ in $2+1$ dimensional class \cite{hinrichsen}. 
In Fig. 4, we display
$\rho(t)$ as a function of $t$ for 
 $p<p_c$ and $p>p_c$ also. As expected, the density of active sites go to zero (absorbing state) for $p>p_c$ while this density saturates
to some asymptotic value signaling the presence of coexistence 
phase for $p<p_c$.

\begin{figure}
 \includegraphics[width=70mm,height=60mm]{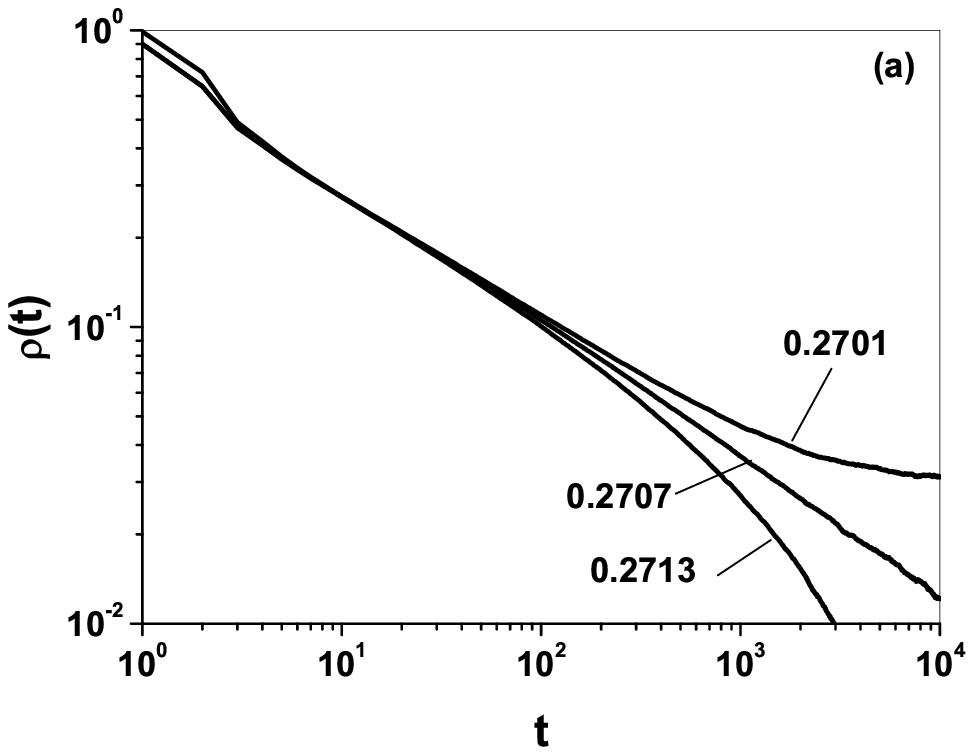}
 \includegraphics[width=70mm,height=60mm]{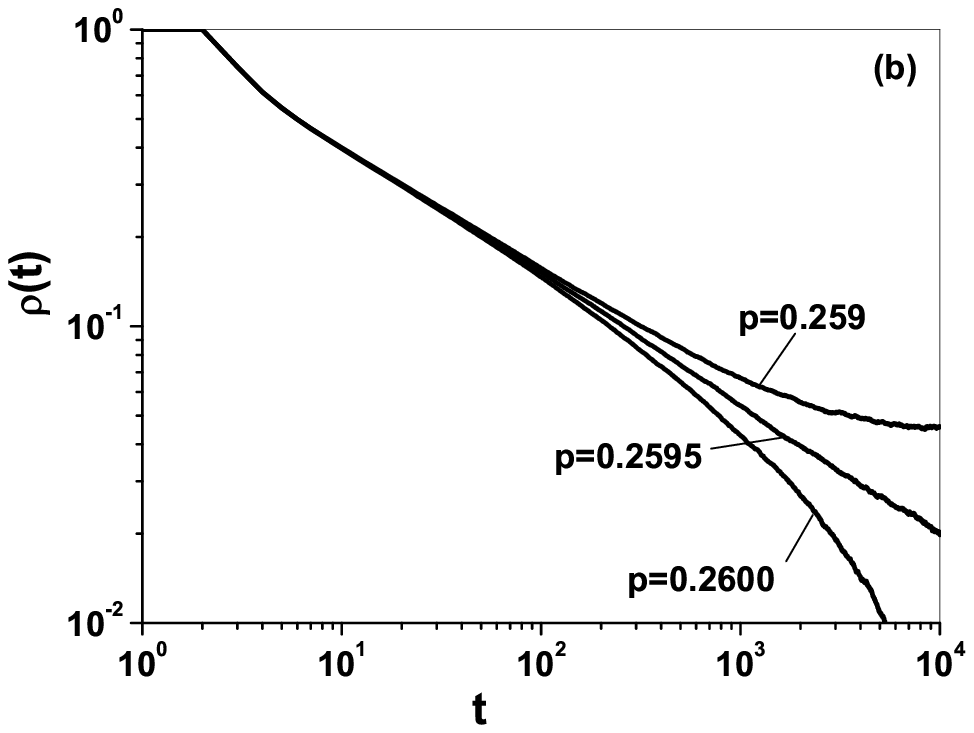}
\caption{\label{fig4}The dynamical behavior of the density of active sites as 
function of time (a) for SPD(P) and  (b) for SPD(T), for lattice size $L=512$. The data averaged over 
$10^3$ samples }
 \end{figure}

In addition, the nonequilibrium phase transitions are characterized by two 
independent correlation length, spatial length scale $\xi_\bot$ and a temporal 
length scale $\xi_\|$. Close to the transition point, these length 
scales are expected to diverge as:
\begin{equation}
\xi_\bot\sim\left|p-p_c\right|^{-\nu_\bot},   \xi_\|\sim\left|p-p_c\right|^{-\nu_\|}
\end{equation}

The two correlation lengths are related by $\xi_\|\propto\xi_\bot^z$
 where $z$ is the dynamic exponent. 
In order to obtain the dynamic exponent  and the two  
correlation exponents, we carry out the off-critical 
simulations and finite size scaling.
For DP, particles density $\rho(t)$ starting from fully 
occupied lattices is expected to scale  
with time and lattice size as follows \cite{hinrichsen}:

\begin{equation}
\rho(t)\sim t^{-\beta/\nu_\|} f(\Delta t^{1/\nu_\|},t^{d/z}/N)
\end{equation}
where $\Delta=\left|p-p_c\right|$ and $N=L^d$ is the total number of 
sites. The exponent 
$\delta$ is given by $\delta=\beta/\nu_\|$.
By plotting the value of $\rho(t) t^\delta$ versus $t\Delta^{\nu_\|}$ for 
different values of $\Delta$  we can 
tune the exponent $\nu_\|$ such that all 
curves collapse on single curve. In Fig. 5, we found the best collapse
is achieved for 
$\nu_\|=1.295 \pm 0.003$ for SPD(P), for the 
SPD(T) $\nu_\|=1.295 \pm 0.004$.

\begin{figure}
\includegraphics[width=70mm,height=60mm]{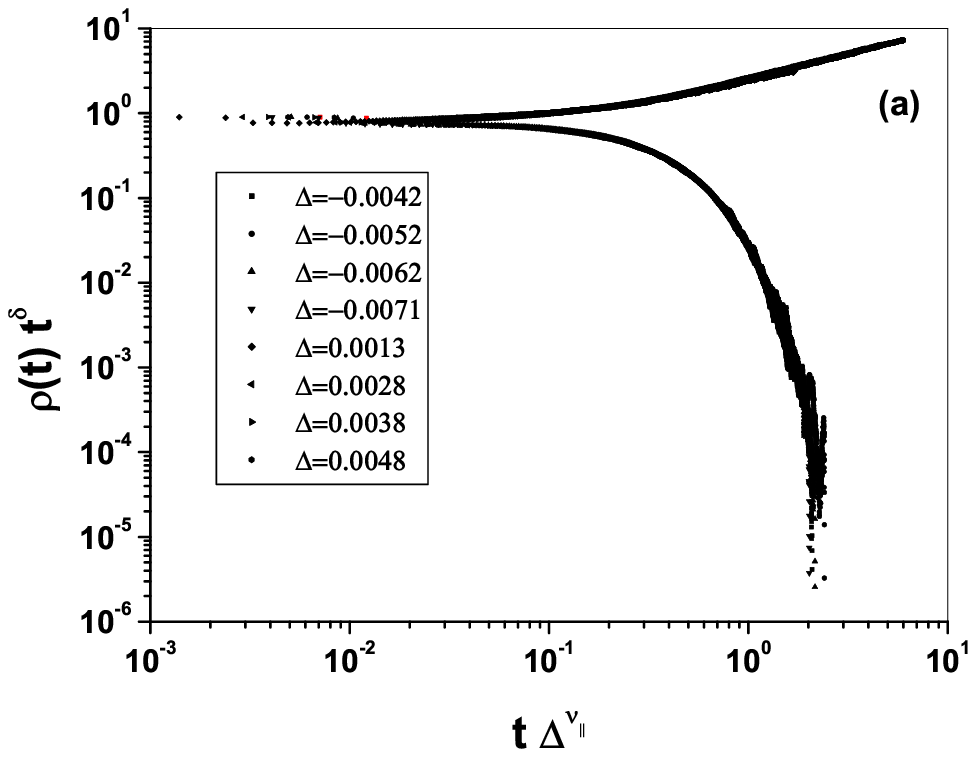}
\includegraphics[width=70mm,height=60mm]{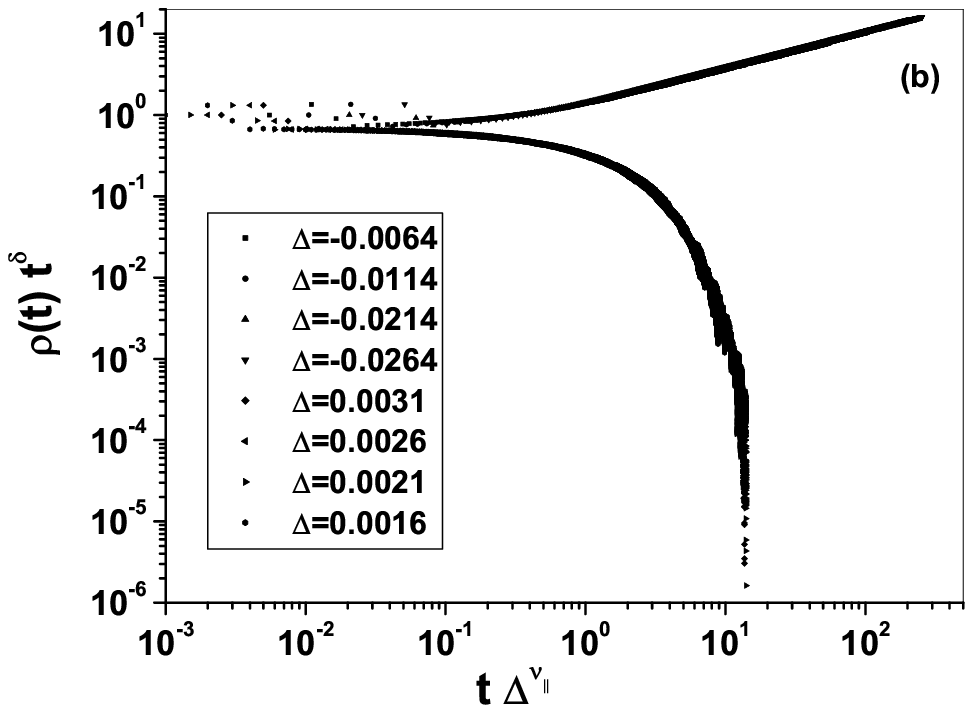}
\caption{\label{fig5}The off-critical scaling function of the density of 
active sites (a) for SPD(P) and (b) for SPD(T). The curves collapse according to the scaling form Eq. (4).}
 \end{figure}

In these simulations, our lattices are  large enough 
so that finite size effects are not very prominent. However, 
as in case of equilibrium scaling, we can carry out finite size
scaling in DP to find further a critical exponents. The system
size enters here as an additional scaling field. 
At the critical point, the finite size simulations can yield us the
value of dynamic exponent $z$ (See Eq. (4)).
We have plotted the  $\rho(t) t^{\delta}$ versus 
$t/N^{\frac{z}{d}}$  for different system size Fig. 6 at $p=p_c$. 
By tuning the value of exponent 
$z$, the best collapse is obtained for  $z=1.76 \pm 0.03$ for  SPD(P)
and $z=1.76 \pm 0.02$ for  SPD(T) which matches
with $z=1.76$ for DP in 2+1 dimensions \cite{hinrichsen}.  
Thus three independent 
exponents $\beta=\delta \nu_{||}$, $\delta$ and $z$ match 
well with DP in 2+1
dimensions for SPD(P) and SPD(T).
Other exponents can be found from these exponents and agree well 
with values in literature. 
For example, the exponent $z$ related to the temporal  and spatial 
correlation
exponents with  that relation $z=\nu_\|/\nu_\bot$. Thus value of  
$\nu_\bot=0.7358$ for both models which matches with value 
quoted in literature.

\begin{figure}
\includegraphics[width=70mm,height=60mm]{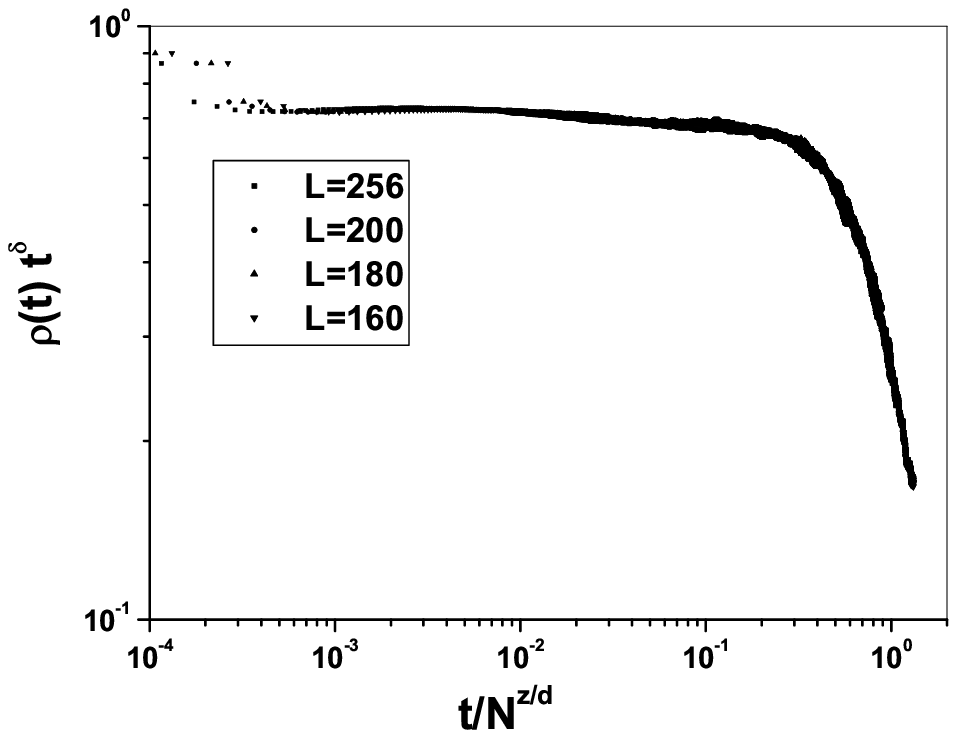}
\includegraphics[width=70mm,height=60mm]{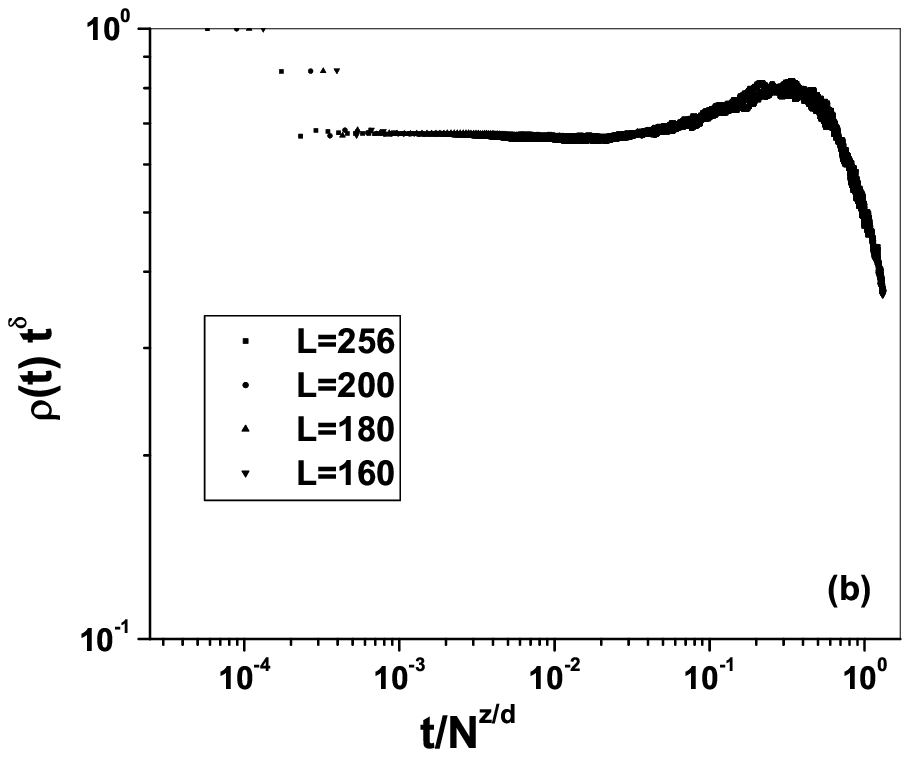}
\caption{\label{fig6}We demonstrate  finite-size scaling 
density of 
active sites at the critical point for various values of lattice sizes 
(a) for SPD(P) and (b) for SPD(T).
An excellent collapse is obtained 
according to the scaling form Eq. (4).}
 \end{figure}
To confirm the results, we make more accurate estimates. These
can be obtained by dynamic simulations starting from a configuration 
which is close to the
absorbing state. (In our system, we cannot start from a single
active site which will disappear immediately.) 
We start our simulation from five active sites are located in the 
center of lattice. We distributed these sites as follows; 
we put one site in the center of lattice and the other four
active sites as the first neighbor of that centered site. In this
case, there is a possibility for the sites in centre to survive 
and grow.
We use the time-dependent 
simulations \cite{grassberger} to estimate values of 
$\theta$ and  $\delta$ (or $z$) and confirm previous results.  
We follow 
the time evolution of this system 
which is initially very close to the
absorbing state \cite{jensen1,dickman}. We numerically measure 
the survival probability $P(t)$ (the probability that the system does not reach 
the absorbing state till time $t$), the average number of active sites $n(t)$, 
and the average mean square distance of spreading of active sites 
from the origin $R^2(t)$. At the critical point these quantities are expected 
to display asymptotic power-laws:
\begin{equation}
P(t)\sim t^{-\delta}
\end{equation}

\begin{equation}
n(t)\sim t^\theta
\end{equation}

and
\begin{equation}
R^2(t)\sim t^{2/z}
\end{equation}

To determine the critical exponents more accurately,
 we adopted the local slope method by introducing 
the {\it{effective exponent}} \cite{hinrichsen,grassberger2}, as follows:
\begin{equation}
-\delta(t)=\frac{\log_{10}(P(t)/P(t/m))}{\log_{10} m}
\end{equation}
where $m$ is a fit parameter. we can get similar definitions 
for the effective exponents
for the quantities $\theta(t)$ an $2/z(t)$. As 
$t\rightarrow \infty$ we should get the right value of the critical exponent.

We use $L=512$ in our simulations and
average over $1.2\times 10^4$ initial conditions. We fix $m=5$.  
In Figs. 7(a), (b) and (c), we show the
values of effective exponents $\delta$, $\theta$ and $2/z$
as a function of $1/t$. For $p\neq p_c$ the values tend to zero or
escape to infinity while they tend to a constant value only
for $p=p_c$. The estimated values 
$\delta=0.434 \pm 0.005 $, $\theta=0.232 \pm 0.004$ and $2/z=1.114 \pm 0.003$ 
are in excellent agreement with the exponents for DP in 2+1 dimensions  
within the error bars.

\begin{figure}
\includegraphics[width=70mm,height=60mm]{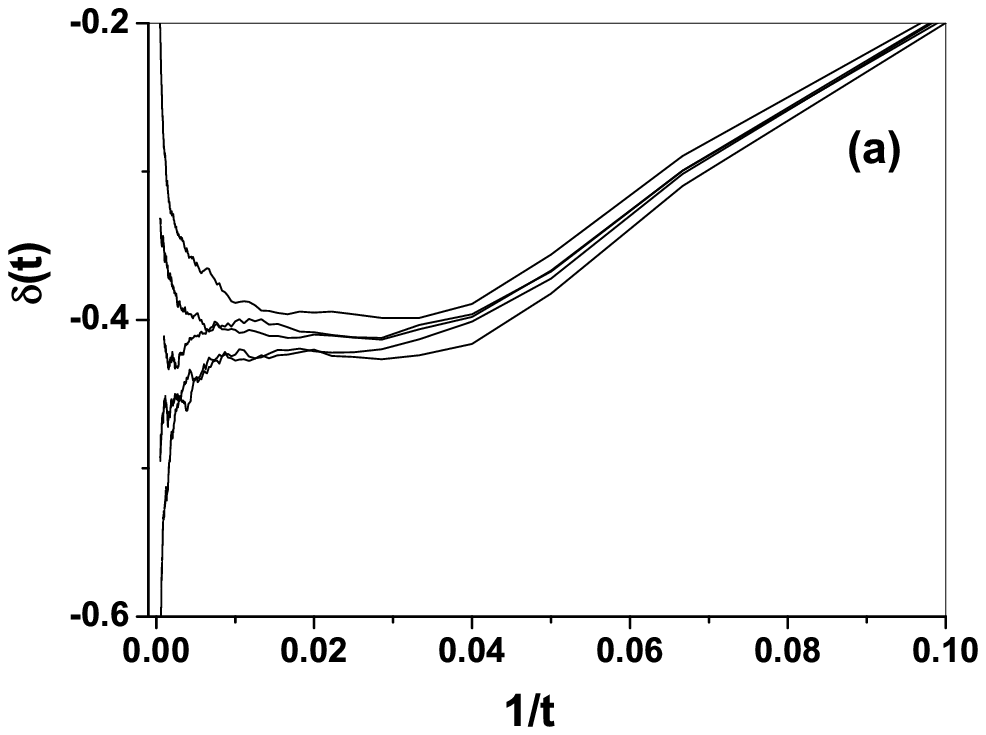}
\includegraphics[width=70mm,height=60mm]{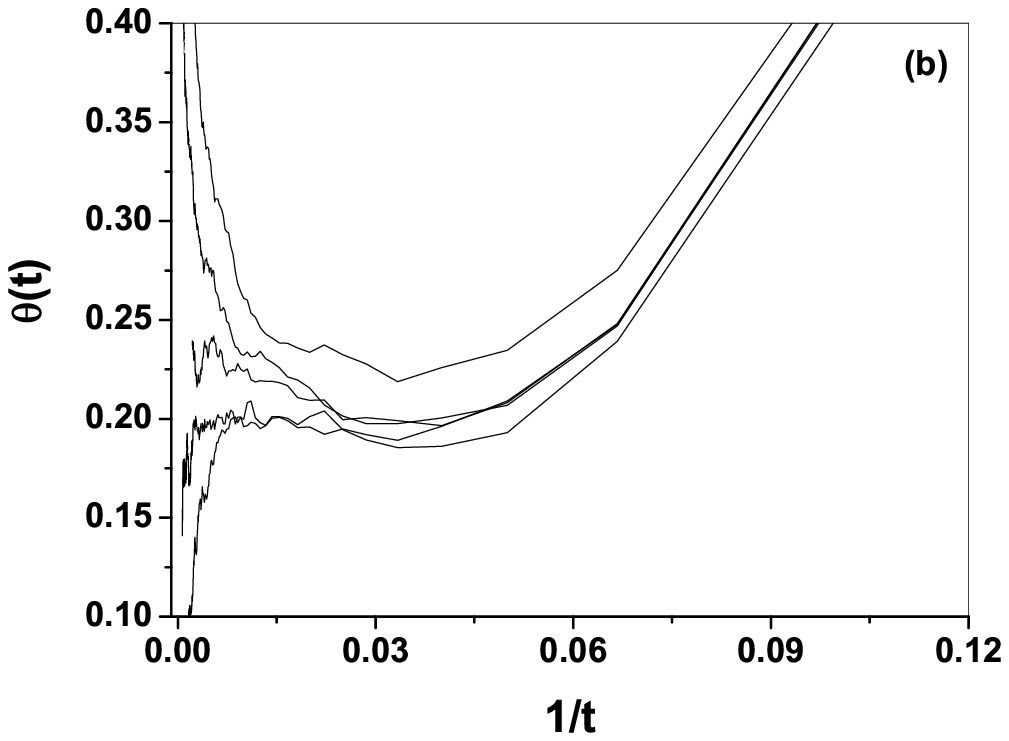}
\includegraphics[width=70mm,height=60mm]{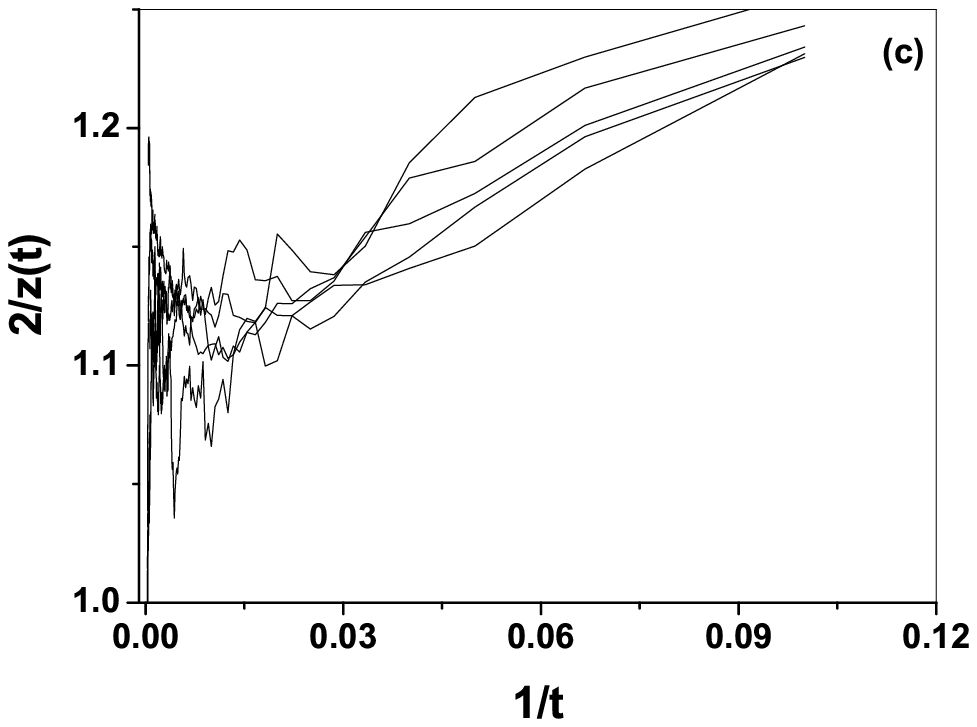}
\caption{\label{fig7}Time dependent behavior of the effective exponents 
(a) $\delta(t)$, (b) $\theta(t)$ and (c) $2/z(t)$ as function of $1/t$ for the 
value of $p=0.2700, 0.2705, 0.2708, 0.2711$ and $0.2714$ 
(from top to bottom curves) for SPD(P).}
 \end{figure}

\subsection{A case of $S<0$}
Some authors believe that the case $S=0$ corresponds
to weak dilemma and we have prisoner's dilemma only for $S<0$. 
We demonstrate that,
our main conclusions remain unchanged for $S<0$.
In Fig. 8, we present the phase diagram for few negative values of $S$, 
namely, 
$S=-0.01, -0.1$ and even $-0.5$ for both models SPD(P) and SPD(T). 
For $S=-0.01$, the phase diagram does not change in  a significant
manner from $S=0$. However,
as one would expect, the area of parameter space which
allows mixed state shrinks with decreasing  values of $S$.
\begin{figure}
 \includegraphics[width=70mm,height=60mm]{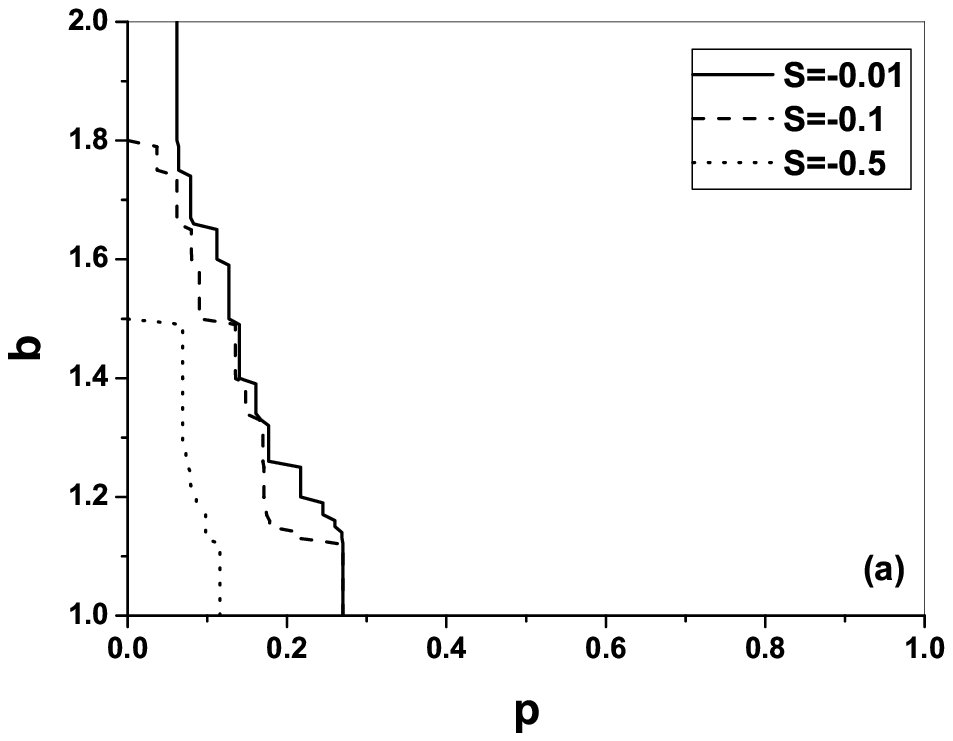}
 \includegraphics[width=70mm,height=60mm]{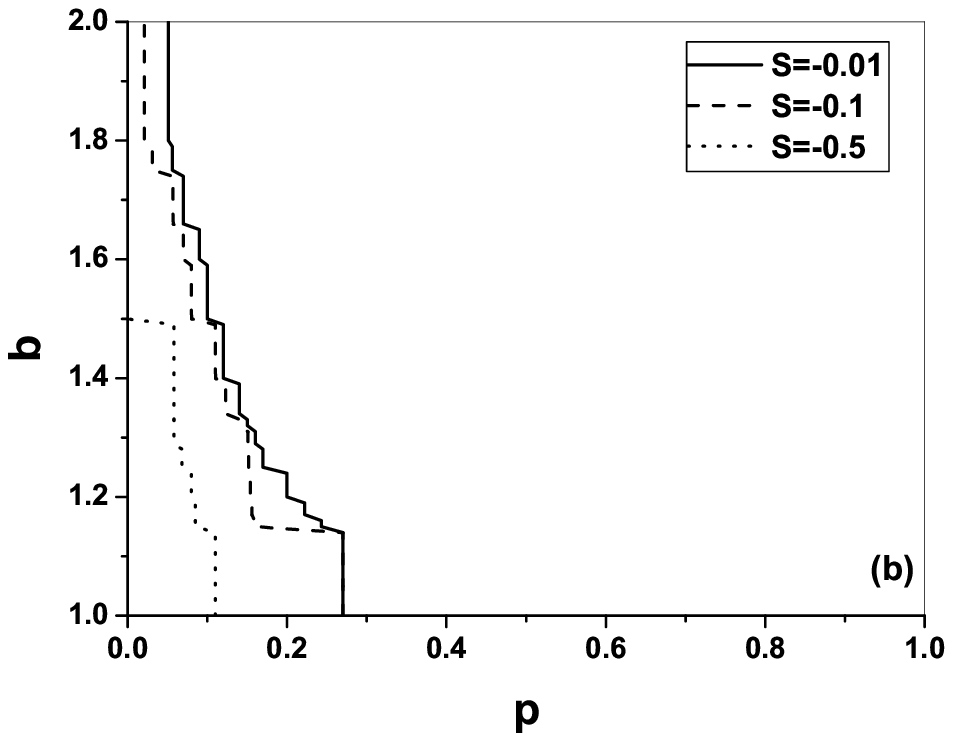}
 \caption{\label{fig8}Schematic phase diagram of (a) SPD(P) and  (b) SPD(T) for $S=-0.01, -0.1$ and $-0.5$. 
The area left to the curves correspond to an active phase whereas the area right to the curves corresponds
to an all-defector phase.
We used lattice size $L=60$ and averaged over $50$ different 
initial samples after discarding $1000$ time-steps }
\end{figure}
  
For $b=1.05$, 
we find the critical parameter value of $p$ for different
values of $S$. For SPD(P), $p_c= 0.268$ for $S=-0.01$ and $p_c= 0.1318$ for
$S=-0.5$. For SPD(T), $p_c= 0.257$ for $S=-0.01$ and $p_c=0.125$ for
$S=-0.5$. We have plotted the density of active sites $\rho(t)$ as a 
function of time at the critical parameter values in Fig. 9.
We clearly see a power law decay of
active sites for $t>1$. The best fit
of the critical exponent for SPD(P) in these cases are 
$\delta=0.432 \pm 0.001$ for $S=-0.5$ and $\delta=0.431 \pm 0.001$ 
for $S=-0.01$. For  SPD(T) $\delta=0.431 \pm 0.002$ when  $S=-0.5$ and
$\delta=0.440 \pm 0.001$ when $S=-0.01$.
These values
match well with the known value of $\delta$ for DP in $2+1$ dimensions.
Thus it is clear that the transition remains in
DP universality class even for negative values of $S$.

\begin{figure}
 \includegraphics[width=70mm,height=60mm]{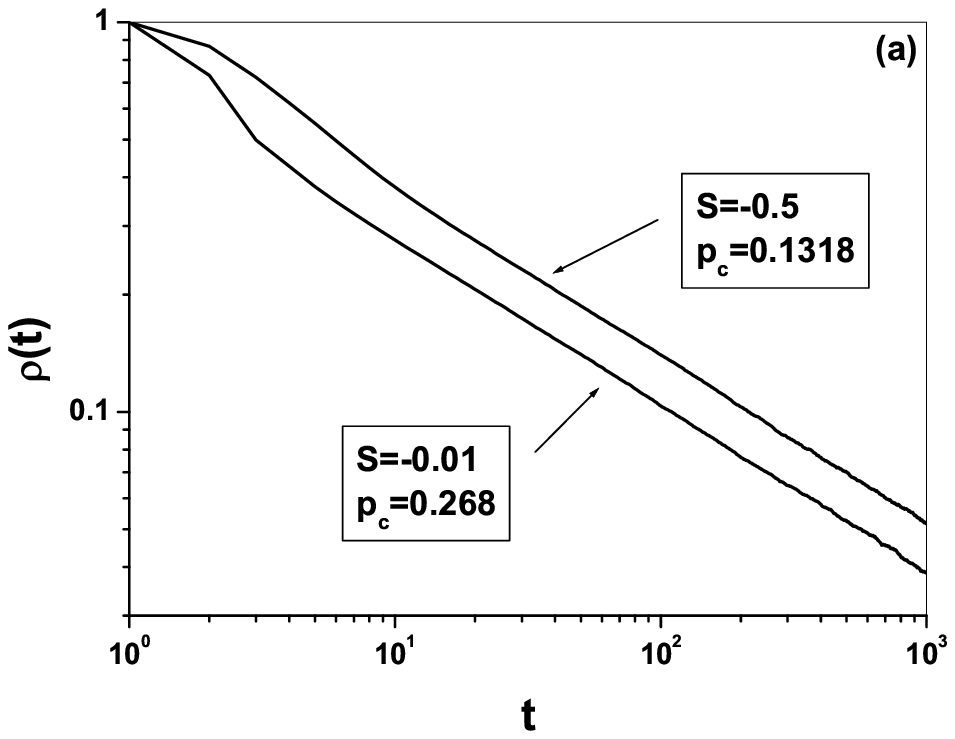}
 \includegraphics[width=70mm,height=60mm]{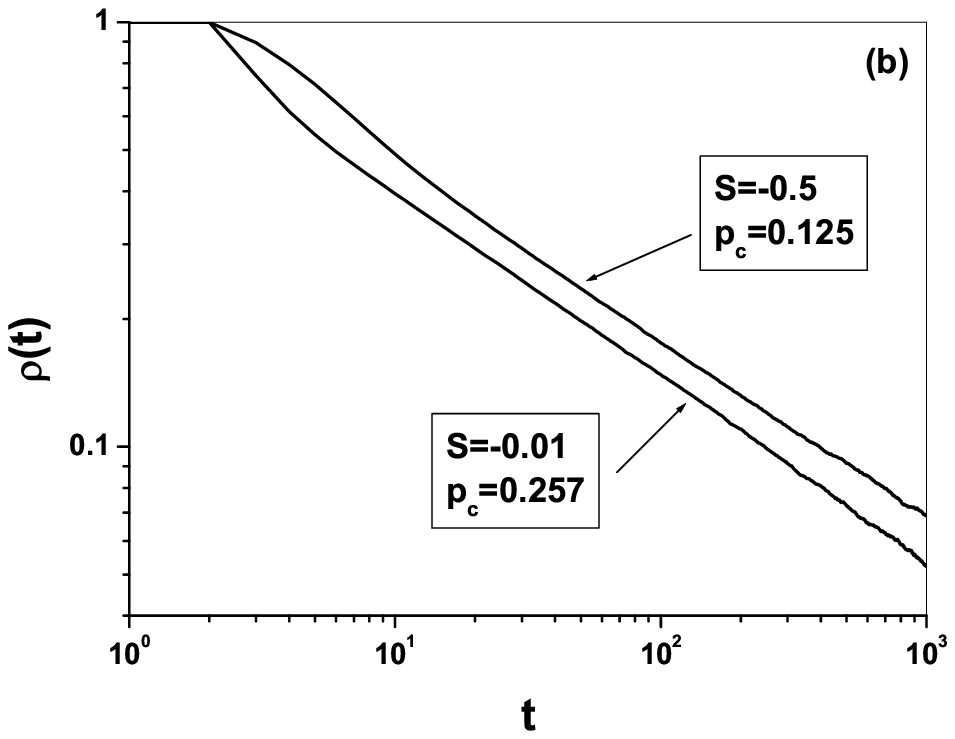}
\caption{\label{fig9}The dynamical behavior of the density of active sites as 
function of time at the critical point when $S=-0.01$ and $S=-0.5$ (a) for SPD(P) and  (b) for SPD(T), for lattice size $L=100$. The data averaged over 
$3\times10^3$ samples.}
 \end{figure}

\subsection{1-D and 3-D case}
PD in 1-D system, when the player $i$ interact with his first two neighbors 
without self-interaction leads to absorbing state for each value of $T>1$. 
Hence, there is no phase transition in this case. However,
preliminary investigation of SPD(P) and SPD(T) in 3-D suggest that, both of 
these model indeed have phase transition which falls in the universality class of 
DP. In Fig.10,  we plot the density of active sites $\rho(t)$ as function of time for both models. 
We use lattice size $L=60$, temptation value $T=1.1$ and each player $i$ interact with his $6$ nearest neighbors 
without self-interaction. At the critical point, the order parameter $\rho(t)$ displays a power-law decay.
 The best fit of the critical exponent is $\delta=0.718\pm 0.008$ for SPD(P) and $\delta=0.723\pm 0.005$ for SPD(T) 
which is in reasonable agreement with the value of $\delta=0.73$ in $3+1$ dimension.  We expect the transition to be in 
DP universality class for higher dimensions as well.
\begin{figure}
 \includegraphics[width=70mm,height=60mm]{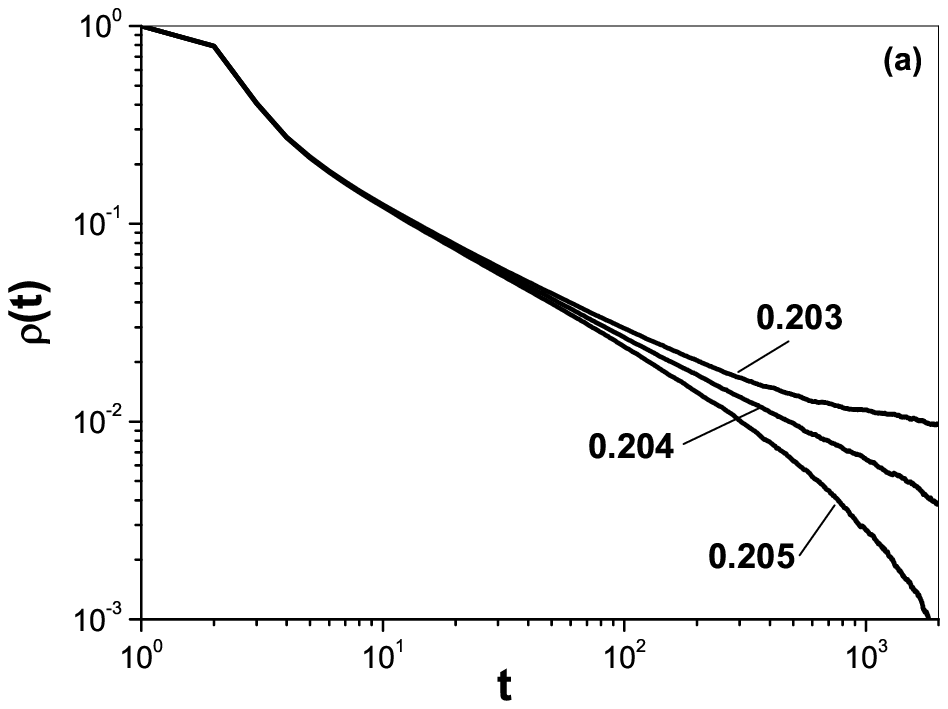}
 \includegraphics[width=70mm,height=60mm]{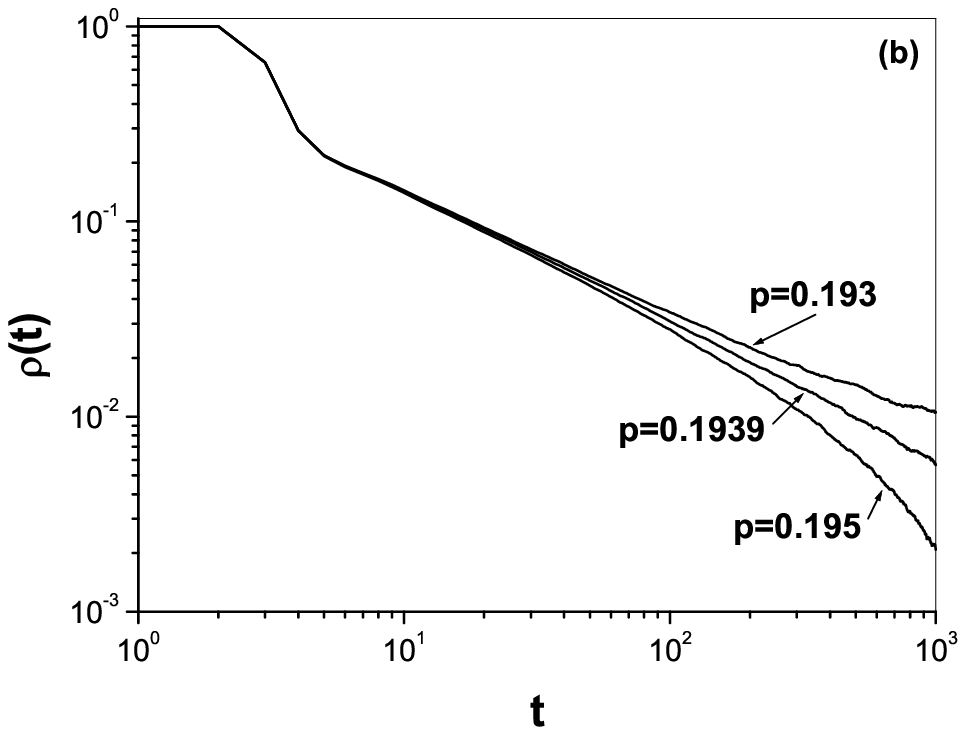}
\caption{\label{fig4}The dynamical behavior of the density of active sites as
function of time in three dimensional case (a) for SPD(P) and  (b) for SPD(T), for lattice size $L=60$. 
The data is averaged over
$200$ samples }
 \end{figure}

\section{Local Persistence}
Recently, persistence has been a fairly popular topic and
has been investigated in great detail in statistical physics.
While most of the studies are
theoretical \cite{Majumdar,Derrida,Krapivsky,
Derrida1,Majumdar1,Lee,Krug,Majumdar2,Kallabis,Chakraborty}, there
have been a few experimental  studies as well \cite{Marcos,Yurke,Tam}. 
It has been shown that, the persistence exponent is a rather nontrivial
quantity to compute even in the simplest of the cases. One needs to know time
correlation at all times and knowing 
it in the asymptotic limit is not good enough. Various
definitions of persistence such as local persistence, global persistence, block
persistence etc. have been proposed \cite{satya}.  Though main object
of studies has been discrete systems, the definition has been suitably modified
and studied also for continuous systems such as coupled maps \cite{Ray,Hu-Gade,Gade-2}.
The most widely studied quantity 
in this context is the local persistence probability $p_l(t)$. It is
 defined as 
the probability that a local variable at a given point of space has not 
changed its state until time $t$ during  stochastic evolution. It 
is observed that, in several systems, at the critical point,
the local persistence probability 
decays algebraically as follows:

\begin{equation}
p_l(t)\sim t^{-\theta_l}
\end{equation}
where $\theta_l$ is the local persistence exponent. This exponent is  found 
to be independent one in the sense that it cannot be obtained 
from other critical exponents. There are no scaling relations to link
it with other exponents. In some cases, different models
displaying continuous transition belonging to same universality
class show the same exponent. For example, the Domany-Kinzel (DK) automata 
in one dimension
and coupled circle maps in one dimension show transition to absorbing
state which is in DP universality class and they show the same 
exponent \cite{Ray}.  However, in general, since the persistence exponent
probes the full evolution of the underlying systems, it may not be
the same in different systems. While we have shown above that the systems
under study unambiguously display a dynamic phase transition in DP universality
class, persistence exponents do not really match the systems studied previously.
 
The definition of persistence has to be appropriately modified 
for absorbing state transitions.
Hinrichsen and Koduvely argued that, the previous definition of local 
persistence not appropriate for the DP class systems. They define the 
local persistence $p_l(t)$ as the probability that inactive site does not 
become active up to time $t$. (The simulations are started from 
random initial conditions.) The reason for this slightly changed
definition is  the asymmetry between active and inactive 
sites in absorbing states models. (The active sites can spontaneously 
turn into inactive sites. Thus number of active sites which do not become 
inactive even once till time $t$ decays exponentially. 
On the other hand, a given inactive site 
may remain inactive for a very long time \cite{hinrichsen3} and will
stay so unless it comes in contact with an active site.) We follow
the same definition for persistence in this work. It is
reasonable in our system since cooperators (active sites) keep defecting
with probability $p$ leading to exponential decay, while defectors 
(inactive sites)
may stay so for really long time.

Local persistence exponent $\theta_l$ in the different DP systems on $1+1$ 
dimension is found to be approximately $\theta_l=1.5$  \cite{hinrichsen3,albano,fuchs,menon}. 
However, in the case of the $2+1$ dimension there is no exact estimate  of the 
value $\theta_l$. 
In the table I, we tabulate  the values of $\theta_l$ for the different 
systems showing DP transition in $2+1$ dimension.\\ 

\begin{tabular}{ccccccc}
\multicolumn{6}{l}{Table I: The  value of $\theta_l$ in  
DP in 2+1 dimensions}\\
\hline
Model& ZGP & CP & CP & Bond-DP& SPD(P)& SPD(T)\\
Ref.& \cite{albano}& \cite{fuchs}&\cite{grass}& \cite{fuchs}& This work& This work\\
\hline
d=2& 1.50(1)& $>$1.62&1.611(1)& $>$1.58& 1.73 $\pm$0.02& 2.24 $\pm 0.03$\\
\hline
\end{tabular}\\

We carried out
simulations at critical point for $L=2000$ and averaged 
over $100$ independent runs. Initial condition consists of $35\%$ defectors 
distributed randomly on the lattice sites. In Fig. 11, 
we clearly observe that the number of persistent sites $p_l(t)$
decays as
a power-law at the critical point for both SPD(P) and SPD(T) models for three decades .
The best power law fit is obtained for $\theta_l=1.73$ for SPD(P) and $\theta_l=2.24$ for SPD(T).
We also plot $p_l(t) t^\theta_l$ as a function of $t$ and the curve is flat for almost three decades
or more. 
We have estimated $\theta_l$ in two more  ways. We carried out extensive simulations for
$L=1024$. We carry out effective exponent analysis and also scaling for off-critical simulations.
The effective exponent analysis is presented in Fig. 12. 
We make a fit as suggested in original work by Grasberger (eq. 9 of ref. \cite{grassberger2} )  at the critical point
and get an estimate of the value in the limit $1/t \rightarrow 0$. 
The fluctuations at large times for persistence (due to smaller data and finite size
effects) are reflected in in the limit $1/t \rightarrow 0$ for effective exponent.
When we make a fit suggested by Grassberger 
for effective exponent,   we obtain $\theta_l=1.72 \pm0.01$ for  SPD(P) and 
$\theta_l=2.25 \pm 0.02$
for SPD(T).

\begin{figure}
\includegraphics[width=70mm,height=60mm]{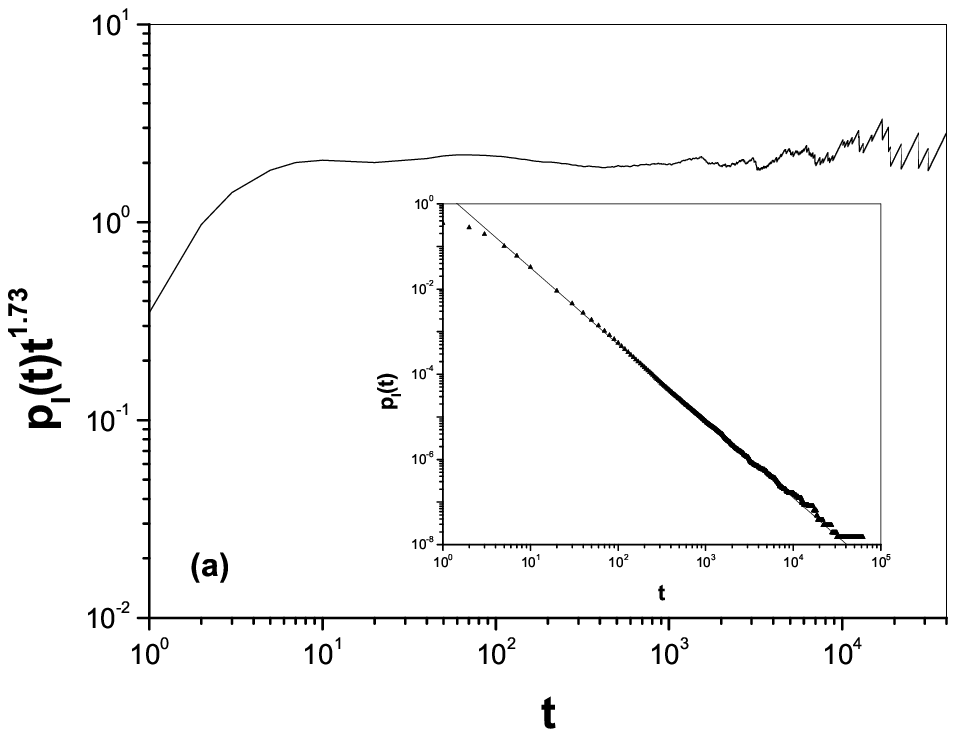} 
\includegraphics[width=70mm,height=60mm]{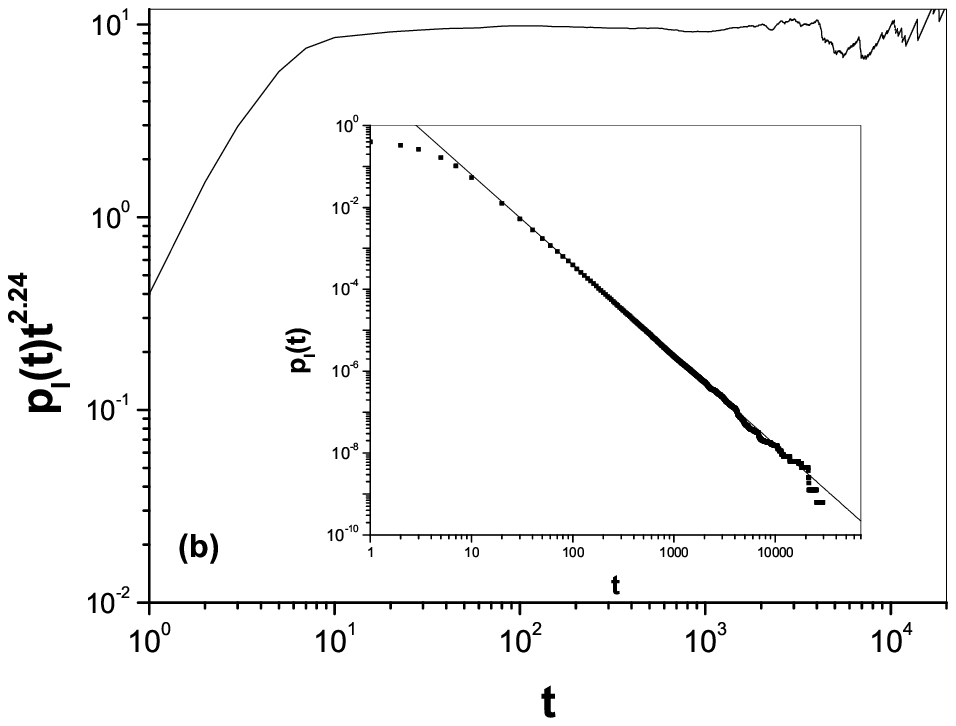} 
\caption{\label{fig10} The local persistence probability 
is plotted as a function of  $t$ for, (a) SPD(P) model at 
 $p_c$. The  
best fit for power-law decay at the critical
point has a slope $\theta_l=1.73$. (b) SPD(T) model at 
 $p_c$. The best fit for power-law at the
critical point  has a slope $\theta_l=2.24$. In both cases, the
lattice size  $L=2000$ and we average over $100$ different
initial conditions.
}
\end{figure} 

\begin{figure}
\includegraphics[width=70mm,height=60mm]{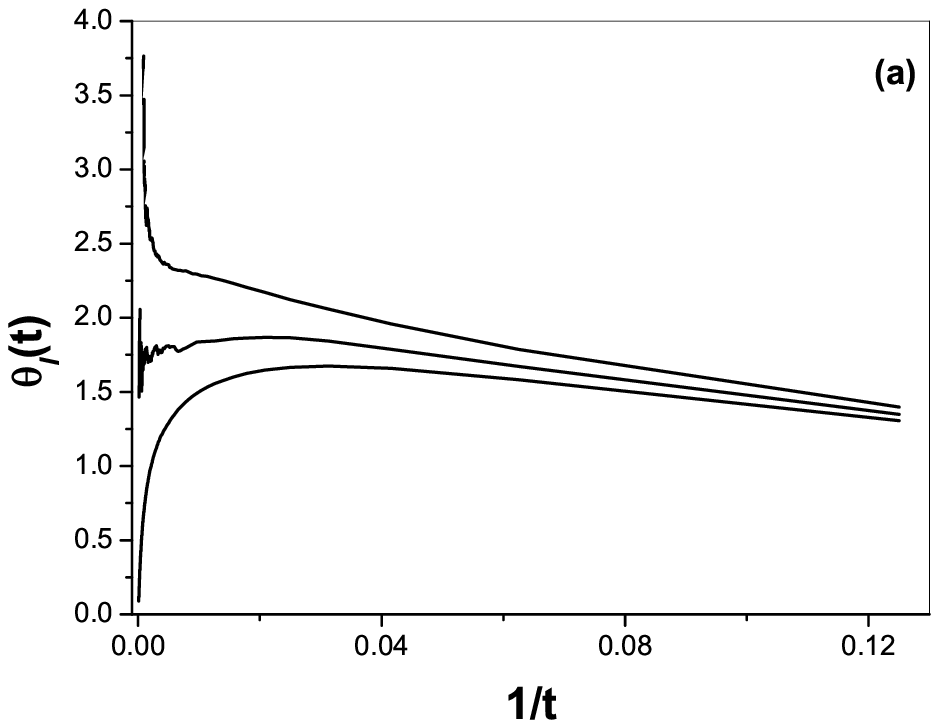} 
\includegraphics[width=70mm,height=60mm]{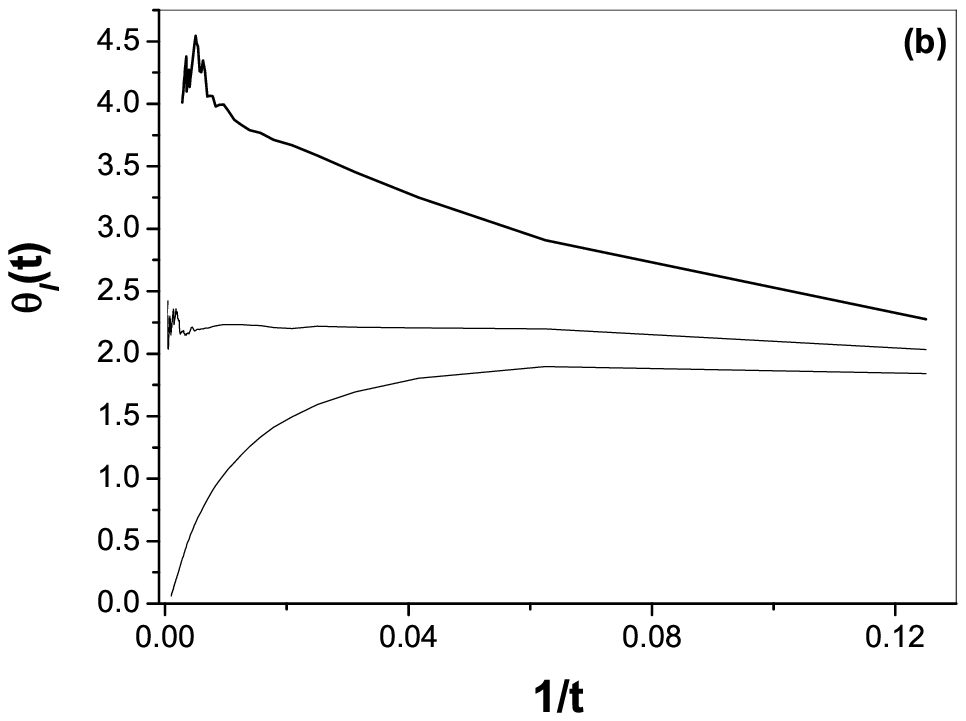} 
\caption{Time dependent behavior of the persistence 
exponents $\theta_l(t)$ as function of $1/t$ (a) for SPD(P) at the 
values of $p=0.2620$, $0.2704$ and $0.2740$  
(from top to bottom curves) (b) for SPD(T) at the 
values of $p=0.2400$, $0.2584$ and $0.2600$  
(from top to bottom curves).}
\end{figure}         
  
The above exponents are confirmed by studying the scaling behavior
of the local persistence 
probability. In analogy to other DP quantities, the local persistence is
expected to have a scaling law of the form \cite{fuchs}:

\begin{equation}
p_l(t) \sim t^{-\theta_l} F(t \Delta^{\nu_\|})
\end{equation}
 where $\Delta=\left|p-p_c\right|$ measures the distance from the critical 
point, F is the off-critical scaling function and $\nu_\|=1.295$ is the 
temporal dynamical exponent of DP. 
 
In the Fig. 13, we have 
plotted the value of $p_l(t) \Delta^{-{\theta_l}{\nu_\|}} $ 
against $t \Delta^{\nu_\|}$ for various values of the parameter $p$. The 
curves shows us a good collapse when the value of $\theta_l = 1.73 \pm 0.02$. Similarly,
for SPD(T), the best collapse is obtained for $\theta_l =2.24 \pm 0.03$. Hence, 
we conclude that the best estimates for persistence exponent are $1.73 \pm 0.02$ for
SPD(P) and $2.24 \pm 0.03$ for SPD(T). 
It could be noted that the values of these exponents 
are much higher than those obtained in 2-d directed percolation in previous studies.

\begin{figure}
\includegraphics[width=70mm,height=60mm]{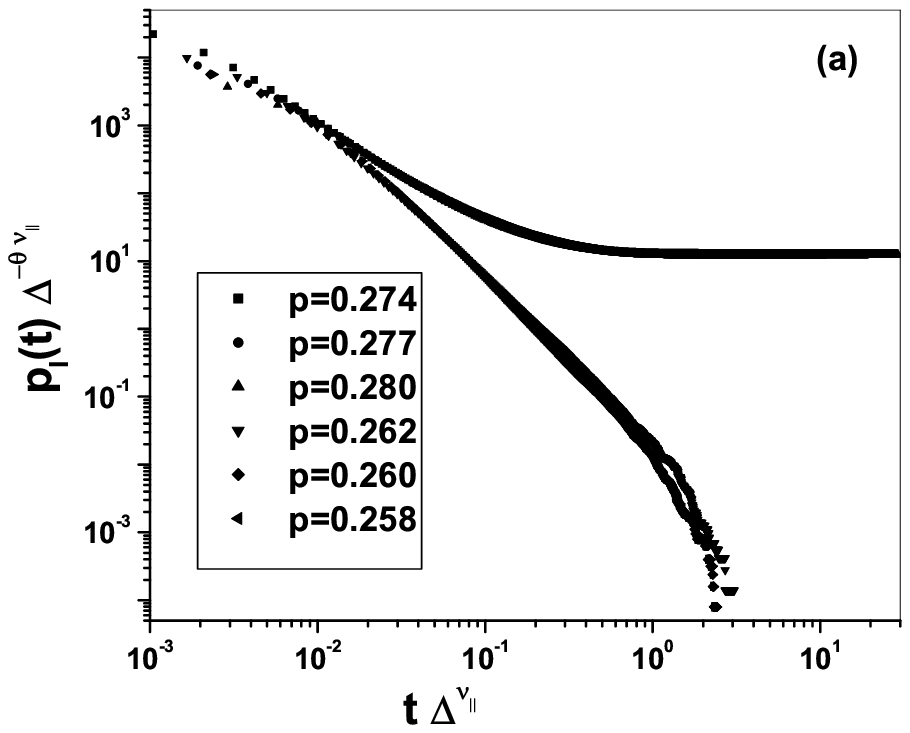}
\includegraphics[width=70mm,height=60mm]{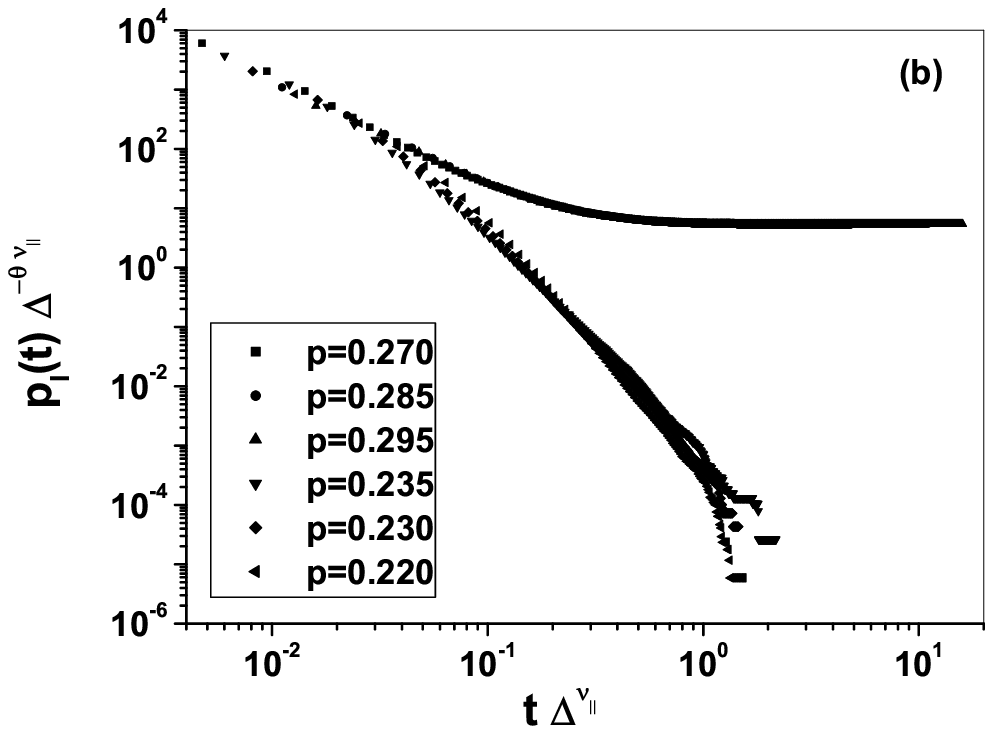}     
\caption{\label{fig11}The off-critical scaling function of the local 
persistence near the critical point for $L=1024$. We average over $100$
different initial conditions. 
(a) SPD(P) model: the best collapse is obtained 
when $\theta_l=1.73$ for SPD(P). 
(b) SPD(T) model: the best collapse is obtained 
when $\theta_l=2.24$.}
\end{figure}

\section{Conclusions}

Good models for realistic situations in ecological and social systems 
require robustness with respect to some degree of noise. The game 
theoretical models with stochastic modifications are relatively less 
studied and in this paper we have tried to investigate two such 
models in detail.  We have studied their phase diagrams and also studied 
the nature of dynamic transition between the two phases observed in these
systems.  In particular, we have studied  two stochastic variants of 
prisoner's dilemma (SPD), (SPD(T) and SPD(P)), on a two dimensional lattice. 
Our investigations  indeed confirm that the results from original model, 
prisoner's dilemma on a lattice, are reasonably robust with respect to noise.
In the models we studied, the
cooperators turn defectors temporarily or permanently. While
SPD(P) was studied previously, SPD(T) is introduced by us in this 
work.
The difference between these models is the following:
In SPD(P) model, the cooperators spontaneously become defectors with
probability $p$ and stay so unless a cooperator in vicinity has higher
payoff. On the other hand, in SPD(T) the defect temporarily for one timestep.
In both the models, depending on value of parameter $p$, the system is found to 
be in the mixed phase or an all-defector phase. The memory is of one timestep
only and the neighboring site does not distinguish between pure defector
and a cooperator who has temporarily turned a defector.
The phase diagrams are studied in detail and for a higher
tendency of cooperators to defect, the mixed systems breaks down 
and we have an all-defector state. 
This is clearly an absorbing state transition.

We have carried out heavy and systematic computation on this system and
a  clear evidence has been presented that both SPD(P) and SPD(T) 
display a transition in the DP universality class. All the DP exponents
have been found and a clean scaling behavior is presented 
in both cases. Of late, persistence in spatially extended dynamical systems has
been a topic of intensive studies in nonequilibrium 
statistical physics. Systems displaying 
phase transition in universality class of directed percolation 
have also been studied in this regard and the persistence exponent
in two dimensions is found to be in the range $1.5-1.6$. The
persistence exponents in our systems are found to be significantly
higher and in one of the cases the exponent is well beyond two. 
This should put any possible speculation about superuniversality
of this exponent to rest. In this case, a clean scaling behavior
is presented which demonstrates the validity of conventional scaling.
To the best of our knowledge such clean scaling
of persistence has not been shown in $2+1$ dimensions. 
As found in spin systems, the 
persistence exponent is the least universal of  critical exponents \cite{Ray}.
For example, let us consider Ising model at finite temperature.
Though the persistence probability shows the same behavior whether 
heat bath algorithm is employed or Glauber algorithm is employed
for temperature $T< T_c$, it is very different for temperatures
$T\geq T_c$, where $T_c$ is the critical temperature \cite{cueille}.
However, it is an interesting fact that this quantity displays
a power law at the critical point. In some cases, the 
exponent has been found theoretically. However,
it is still a puzzle
what this exponent means physically. For example, it is not
clear which new physical insight is brought by having the information that 
the persistence exponent 0.1207  for 1-D diffusion, a problem which
is fully understood and is exactly solvable \cite{satya}.

There are several other variants which could be an 
object of studies in future. If we consider cooperation with probability
$p$
and defection as two possible strategies and if we impose a condition that
the strategies are mimicked and not behavior in previous
time step, this variant can have two possible
absorbing state and an interesting phase diagram. We will
be investigating this variant in future studies. One could also
consider dynamic phase transitions
in presence of random introduction of defectors and cooperation as
well as effect of asynchronicity. Nature of dynamic phase transitions
in game-theoretic systems is a rich and unexplored field and
it could yield interesting insights. In this work, we have 
brought out two  models which are unambiguously in the
universality class of directed percolation. Though their
critical exponents match with standard DP exponents in $2+1$ dimensions,
they have
widely different persistence exponents. Thus having same exponent
in two different systems as in coupled map lattice in one dimension
and DK automata as pointed out by Menon {\it{et al.}} \cite{menon}
could be a coincidence or presence of
certain dynamical properties which needs further investigation.

\section{Acknowledgment}
MAS thanks Govt. of Yemen for scholarship
and PMG thanks DST for financial support.

\end{document}